\documentclass[twocolumn,showpacs,aps,prd,superscriptaddress]{revtex4}

\usepackage{graphicx}
\usepackage{dcolumn}
\usepackage{amsmath}
\usepackage{epsfig}

\usepackage{enumitem}

\input babarsym

\begin{document}

\begin{flushleft}
\babar-PUB-16/015\\
SLAC-PUB-16822\\[5mm]
\end{flushleft}

\title{
{\large \bf
Measurement of the $\Bz\to\Dstarm\pip\pim\pip$ branching fraction
} 
}

%
%
\author{J.~P.~Lees}
\author{V.~Poireau}
\author{V.~Tisserand}
\affiliation{Laboratoire d'Annecy-le-Vieux de Physique des Particules (LAPP), Universit\'e de Savoie, CNRS/IN2P3,  F-74941 Annecy-Le-Vieux, France}
\author{E.~Grauges}
\affiliation{Universitat de Barcelona, Facultat de Fisica, Departament ECM, E-08028 Barcelona, Spain }
\author{A.~Palano}
\affiliation{INFN Sezione di Bari and Dipartimento di Fisica, Universit\`a di Bari, I-70126 Bari, Italy }
\author{G.~Eigen}
\affiliation{University of Bergen, Institute of Physics, N-5007 Bergen, Norway }
\author{D.~N.~Brown}
\author{Yu.~G.~Kolomensky}
\affiliation{Lawrence Berkeley National Laboratory and University of California, Berkeley, California 94720, USA }
\author{H.~Koch}
\author{T.~Schroeder}
\affiliation{Ruhr Universit\"at Bochum, Institut f\"ur Experimentalphysik 1, D-44780 Bochum, Germany }
% ...
\collaboration{The \babar\ Collaboration}
\noaffiliation

\begin{abstract}
Using a sample of ${(470.9 \pm 2.8) \times 10^6 ~ B \kern 0.2em\overline{\kern -0.2em B}}$ pairs, we measure the decay branching fraction ${\mathcal{B}(\B^0 \to D^{*-} \pi^+ \pi^- \pi^+) = (7.26 \pm 0.11 \pm 0.31) \times 10^{-3}}$, where the first uncertainty is statistical and the second is systematic. Our measurement will be helpful in studies of lepton universality by measuring ${\mathcal{B}(\B^0 \to D^{*-} \tau^+ \nu_\tau)}$ using ${\tau^+ \to \pi^+ \pi^- \pi^+ \overline{\nu}_\tau}$ decays, normalized to ${\mathcal{B}(\B^0 \to D^{*-} \pi^+ \pi^- \pi^+)}$.
\end{abstract}


\maketitle

The \babar\ Collaboration measured the branching fraction ratios for $B$ semileptonic decays to $D$
  and $D^*$
  
\begin{equation}
\mathcal{R^{(*)}} = \frac{\mathcal{B}(\Bbar \to D^{(*)} \taum \nutb)} {\mathcal{B}(\Bbar \to D^{(*)} \ell^- \nub_l)} ,
\end{equation}
\noindent where $\ell^-$ is an electron or a muon, to be in excess of standard model (SM) predictions~\cite{ref:babarDstartaunu}. The use of charge conjugate reactions is implied throughout this article. After combining the results for $\mathcal{R}$ and $\mathcal{R^{*}}$, the excess is inconsistent with lepton universality at the 3.4$\sigma$ level. The Belle Collaboration~{\cite{ref:Belle} and the LHCb Collaboration~{\cite{ref:LHCb}} conducted similar measurements with comparable results. A measurement of ${\mathcal{B}(\Bz\to\Dstarm\taup\nut)}$ using ${\taup \to \pip\pim\pip \nutb}$ decays, normalized to ${\mathcal{B}(\Bz\to\Dstarm\pip\pim\pip})$, may yield the observation of a further deviation from the SM. Such a measurement has not been done before and may make use of a clean kinematic signature. This possibility relies in part on a measurement of ${\mathcal{B}(\Bz\to\Dstarm\pip\pim\pip})$, for which the current world average value is $(7.0\pm0.8)\times10^{-3}$~\cite{ref:pdg}. The LHCb Collaboration measured this value to be $(7.27~\pm~0.11$(stat.)$~\pm~0.36$(syst.)$~\pm~0.34$(norm.))$ \times 10^{-3}$~\cite{ref:LHCb2}, where the final uncertainty is due to using $\Bz\to\Dstarm\pip$ decays for normalization purposes. This measurement has not been included in the world average value as yet. In this article, we report on a measurement of ${\mathcal{B}(\Bz\to\Dstarm\pip\pim\pip)}$.

We use data recorded with the \babar\ detector at the \pep2\ asymmetric-energy \epem\ collider at SLAC. The \babar\ detector is described in detail elsewhere~\cite{ref:detector,ref:detector2}. The data sample corresponds to an integrated luminosity of $424.2\pm1.8$ \invfb collected at the \FourS resonance~\cite{Lees:2013rw}, which corresponds to the production of $(470.9 \pm 2.8) \times 10^6$ \BB pairs. We use Monte Carlo (MC) simulations to understand background processes and signal reconstruction efficiencies. The EvtGen event generator~\cite{ref:evtgen} is used to simulate particle decays. This includes a sample of $\epem\to\qqbar (\gamma)$ events, where $q$ is a $u, d, s,$ or $c$ quark, with an equivalent luminosity of 2,589\invfb and a sample of $1,427 \times 10^6$ \BB pairs. The detector response is simulated with the Geant4~\cite{ref:geant4} suite of programs.

We fully reconstruct the {\Bz\to\Dstarm\pip\pim\pip} decay chain by adding the four-momenta of particle candidates. The \Dstarm mesons are reconstructed in the $\Dstarm\to\Dzb\pim$ and $\Dzb\to\Kp\pim$ final states. A \Dzb candidate is reconstructed from two charged-particle tracks, of which one is identified as a \Kp meson based on information obtained using the tracking and Cherenkov detectors. We require \Dzb candidates to have an invariant-mass value within $\pm20\mevcc$ of the nominal \Dzb mass~\cite{ref:pdg}, which corresponds to 3 standard deviations in its mass resolution. Each \Dzb candidate is combined with a charged-particle track with momentum less than 0.45\gevc in the \epem center-of-mass (CM) frame to form a \Dstarm candidate. We require the difference between the reconstructed mass of the \Dstarm candidate and the reconstructed mass of the \Dzb candidate to lie between 0.1435 and 0.1475\gevcc. The \Dstarm candidate is combined with three other charged-particle tracks to form a \Bz candidate. We do not explicitly apply particle identification to select charged pions, but assign the pion mass hypothesis to all tracks other than the \Kp daughter 
of the \Dzb. All other reconstructed tracks and neutral clusters in the event are collectively referred to as the rest of the event (ROE). We use a neural network classifier \cite{ref:TMVA} to suppress non-\BB backgrounds. The classifier makes use of nine variables, each of which is calculated in the CM frame:

\begin{itemize}

\item the cosine of the angle between the \Bz candidate's thrust axis \cite{ref:thrust} and the beam axis;
\item the sphericity \cite{ref:spher} of the \Bz candidate;
\item the thrust of the ROE;
\item the sum over the ROE of $p$, where $p$ is the magnitude of a particle's momentum;
\item the sum over the ROE of $\frac{1}{2}(3\cos^2 \theta - 1) p$, where $\theta$ is the polar angle of a particle's momentum;
\item the cosine of the angle between the thrust axis of the \Bz candidate and the thrust axis of the ROE;
\item the cosine of the angle between the sphericity axis of the \Bz candidate and the thrust axis of the ROE;
\item the ratio of the second-order to zeroth-order Fox-Wolfram moment using all reconstructed particles \cite{ref:r2all};
\item the cosine of the angle between the thrust axis calculated using all reconstructed particles and the beam axis.

\end{itemize}

\noindent Each of these nine variables contributes to separating \Bz decays from non-\BB decays. We apply a selection on the output of the neural network classifier that rejects 69\% of reconstructed signal candidates from non-\BB decays, and retains 80\% of correctly reconstructed \Bz candidates. Finally, we require the \Bz candidate to have a CM frame energy within $\pm90\mev$ of $\sqrt{s}/2$, where $\sqrt{s}$ is the nominal invariant-mass of the initial state. This corresponds to 4 standard deviations in the energy resolution. We retain all \Bz candidates that pass our selection criteria instead of selecting a best candidate for each event. In MC-simulated signal and background events that have at least one \Bz candidate passing all selection criteria, there are on average 1.57 and 1.37 \Bz candidates per event, respectively.

After applying all selection criteria, we determine the energy-substituted mass ${m_{\rm ES} = \sqrt{s/4 - p_B^2}}$ for the selected \Bz candidates, where $p_B$ is the CM-frame momentum of a \Bz. Figure \ref{fig:mES} shows the $m_{\rm ES}$ distribution for the data and for MC-simulated events. The $m_{\rm ES}$ distribution of correctly reconstructed signal candidates has a peak near the \Bz mass.

\begin{figure}
\includegraphics[width=\columnwidth]{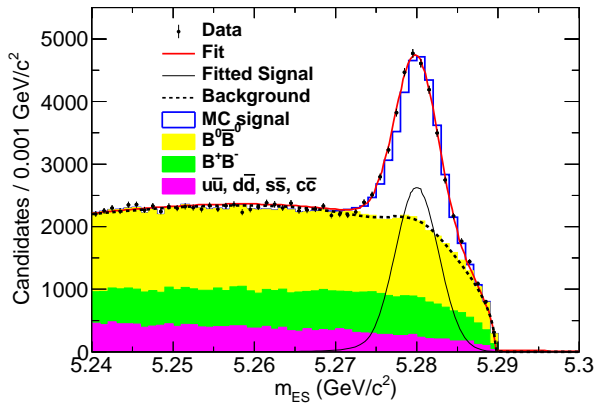}
\caption{
(color online) The $m_{\rm ES}$ distribution of \Bz candidates for data (points), MC simulations (histograms), and the unbinned extended-maximum-likelihood fit to the data (curves). The MC distributions are shown as stacked histograms. The $\Bz\to\Dstarm D_s^+$ with $D_s^+ \to \pip\pim\pip$ decays are part of the MC signal. The MC signal contribution is normalized such that its stacked histogram has the same integral as the data. The components of the MC simulations and the fit are described in the legend. The $m_{\rm ES}$ peak of the MC signal is slightly above that of the data. This shift has a negligible effect on the signal yield.
}
\label{fig:mES}
\end{figure}

The $m_{\rm ES}$ distribution of signal events is modeled using a Crystal Ball~\cite{ref:crys} probability density function (PDF), with cutoff and power-law parameters determined using MC-simulated events.  We consider only \Bz candidates that are correctly reconstructed. We model the background $m_{\rm ES}$ distribution as follows. The non-peaking backgrounds from $\epem\to\qqbar (\gamma)$ events and from \BB pairs are modeled using an ARGUS function~\cite{ref:argus}. Each of the peaking backgrounds from \BpBm and \BzBzb is modeled by a Gaussian distribution for which the normalization, mean, and width, are determined by a fit to the corresponding simulated event sample. We perform a one-dimensional unbinned extended-maximum-likelihood fit in order to estimate the number of signal candidates. We allow the mean and width parameters of the Crystal Ball function, the curvature parameter of the ARGUS function, and the normalization of the non-peaking background, to vary in the fit. The cutoff parameter for the ARGUS function is fixed to $\sqrt{s}/2$, and the peaking background PDF shapes and normalizations are fixed to their MC-estimated values. The peaking backgrounds contributions are estimated to be $590\pm120$ and $1450\pm130$ candidates from \BpBm and \BzBzb decays, respectively; some originate from signal decays where one or more pion is misreconstructed even when there is a correctly reconstructed \Bz candidate. There is also a contribution from $\Bp\to\Dstarm X$ and $\Bz\to\Dstarm X$ decays, where $X$ denotes any combination of $\pi$ and $\rho$ mesons other than $\rho^0 \pip$ or $\pip\pim\pip$. The fit to the $m_{\rm ES}$ distribution shown in Fig.~\ref{fig:mES} results in a signal yield of $17800\pm300$.

The distribution for the MC signal peaks at a higher $m_{\rm ES}$ value than the data. We repeat the fit procedure on our MC sample where we correct for this difference. The effect on the signal yield is negligible.

We define the signal region to be ${5.273 < m_{\rm ES} < 5.285~\gevcc}$, and a sideband region to be ${5.240 < m_{\rm ES} < 5.270~\gevcc}$. About 97.6\% of signal events are contained within the signal region. To obtain the $3\pi$ invariant mass distribution for the signal events in Fig.~\ref{fig:a_1CMass}, we subtract the events in the sideband region of the $m_{\rm ES}$ in Fig.~\ref{fig:mES}, normalized to the fitted background component in the signal region, from the total $3\pi$ mass distribution. By integrating the dashed line in Fig.~\ref{fig:mES}, we obtain $68883$ events in the sideband-region and $24427$ background events in the signal region. These values make use of the peaking background estimates described in the previous paragraph.

\begin{figure}
\includegraphics[width=\columnwidth]{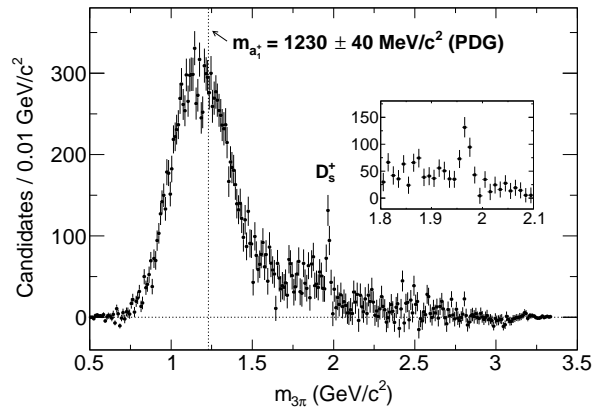}
\caption{
The background-subtracted invariant-mass spectrum of the $3\pi$ system. The indicated mass value of the $a_1^+$ is obtained from Ref.~\cite{ref:pdg}. The $\Bz\to\Dstarm D_s^+, D_s^+ \to \pip\pim\pip$ decay, which is removed in the final result, is visible in the spectrum. The spectrum is obtained prior to the efficiency correction. The inset shows the distribution around the $D_s^+$ region.
}
\label{fig:a_1CMass}
\end{figure}

As expected from the branching fractions in Ref.~\cite{ref:pdg}, the main contribution comes from $a_{1}^+(1260)$ decays, and a contribution from the decay $D_{s}^{+} \to \pip\pim\pip$  is also apparent. There is as well activity in the 1.7--1.9~\gevcc region, which may be due to the $J^P = 0^-$ $\pi (1800)$ meson. The analysis of the $a_1^+$ region is complicated and will be the subject of a separate study.

The $D_s^+$ events result from the doubly-charmed decay $\Bz \to \Dstarm D_s^+$ in which the $D_s^+$ decays weakly to \pip\pim\pip. Since the $D_s^+$ decay results from an entirely different \Bz decay mode, it represents a contamination of our $\Dstarm\pip\pim\pip$ sample. We remove the $D_s^+$ contribution by subtracting the events in the 1.9--2.0~\gevcc region of the $3\pi$ invariant-mass distribution of Fig.~\ref{fig:a_1CMass} that exceed the interpolation of the bin contents in the 1.8--1.9~\gevcc and 2.0--2.1~\gevcc regions. The removed $D_s^+$ contribution amounts to $233\pm63$ events, and the remaining events in the 1.9--2.0~\gevcc region total $326\pm35$.

We estimate the reconstruction efficiency as a function of $3\pi$ invariant-mass using MC-simulated events. This is shown in Fig.~\ref{fig:efficiency}. Since we model the $m_{\rm ES}$ PDF of the signal only considering \Bz candidates that are correctly reconstructed, we apply exactly the same procedure of determining the signal yield in our study of the reconstruction efficiency in order to determine the branching fraction correctly. The efficiency of the decay channel $\Dstarm a_1^+$, where the $a_1^+$ decays to $\rho^0\pip$ and the $\rho^0$ to $\pip\pim$ was studied. The simulation assumes a mass of 1.230~\gevcc and a width of 400~MeV for the $a_1^+$~\cite{ref:pdg}. The reconstruction efficiencies of $\Bz\to\Dstarm\rho^0\pip$ and $\Bz\to\Dstarm D_s^+$ decays are consistent with $\Bz\to\Dstarm a_1^+$ decays. Taking into account the efficiency as a function of the $3\pi$ mass, and removing the $D_s^+$ background, the total number of produced $\Bz\to\Dstarm\pip\pim\pip$ events is estimated to be $84400
\pm1200$.

\begin{figure}
\includegraphics[width=\columnwidth]{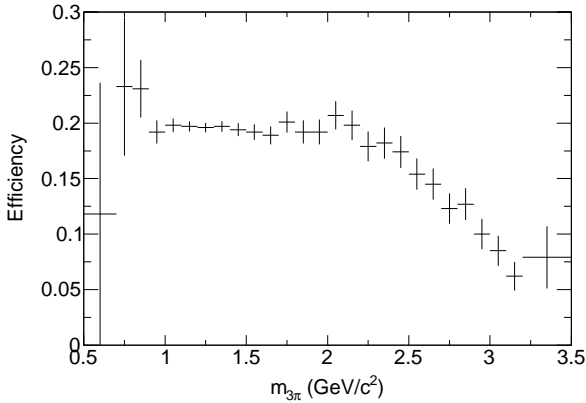}
\caption{
The reconstruction efficiency as a function of $3\pi$ invariant-mass using MC-simulated events. The uncertainties are statistical.
}
\label{fig:efficiency}
\end{figure}

\begin{table}
  \begin{center}
  \caption{Summary of systematic uncertainties. The uncertainties are assumed to be uncorrelated, and so are added in quadrature.}
    \begin{tabular}{l c}
      \hline
      Source &Uncertainty (\%)\\
      \hline
      Fit algorithm and peaking backgrounds&2.4\\
      Track-finding & 2.0\\
      $\pip\pim\pip$ invariant-mass modeling&1.7\\
      \Dstarm and \Dzb decay branching fractions&1.3\\
      $\FourS \to \BzBzb$ decay branching fraction&1.2\\
      \Kp identification & 1.1\\
      Signal efficiency MC statistics&0.9\\
      Sideband subtraction&0.7\\
      $\BB$ counting&0.6\\
      \hline
      Total&4.3\\
      \hline
    \end{tabular}
      \label{tab:uncertainties}
        \end{center}
\end{table} 

\begin{table*}
  \begin{center}
  \caption{Fit parameters obtained from MC-simulated events. These parameters are fixed to the central values in the signal extraction procedure. We perform a toy study where we simultaneously vary these by the quoted uncertainties (along with their correlations, which are not shown in the table) to study systematic effects on the signal yield.}
    \begin{tabular}{l c}
      \hline
      Parameter & Value \\
      \hline
      \BpBm peaking background $m_{\rm ES}$ Gaussian mean & $5.2796\pm0.0006$~\gevcc\\
      \BpBm peaking background $m_{\rm ES}$ Gaussian width & $0.0036\pm0.0003$~\gevcc\\
      Number of \BpBm peaking background & $590\pm120$ \\
      \BzBzb peaking background $m_{\rm ES}$ Gaussian mean & $5.2806\pm0.0002$~\gevcc\\
      \BzBzb peaking background $m_{\rm ES}$ Gaussian width & $0.0029\pm0.0002$~\gevcc\\
      Number of \BzBzb peaking background & $1450\pm130$ \\
      Signal's Crystal Ball PDF cutoff value& $2.09\pm0.08$ \\
      Signal's Crystal Ball PDF power-law value& $3.7\pm0.5$ \\
      \hline
    \end{tabular}
      \label{tab:fitParametersStudy}
        \end{center}
\end{table*} 

Table \ref{tab:uncertainties} summarizes the systematic uncertainties for this analysis. The uncertainties of our extended-maximum-likelihood fit algorithm and peaking backgrounds are estimated together by taking into account the uncertainties of the fixed parameters in the fit. The values we used are shown in Table \ref{tab:fitParametersStudy}. These values are obtained entirely from studies of MC-simulated background samples. Therefore, we consider varying the mean and width of the $m_{\rm ES}$ distributions for the peaking \BpBm and \BzBzb backgrounds, the number of \BzBzb and \BpBm peaking background events, and the Crystal Ball PDF cutoff and power-law parameter values for the signal. These values are sampled from an eight-dimensional Gaussian function with means, widths, and correlations that correspond to the fit results for the PDF's for signal and peaking backgrounds simulated events. The systematic uncertainty is taken as the standard deviation of the distribution of the number of signal events from an ensemble of fits, and is found to be 2.4\%. The systematic uncertainty due to track finding consists of two components: 1.54\% for laboratory momenta less than 0.18~\gevc, a region dominated by tracks from the decay $\Dstarm\to\Dzb \pim$, and 0.26\% for greater than this value~\cite{ref:tracking}. The two components are added in quadrature. The pion from the $\Dstarm\to\Dzb \pim$ decay has momentum less than 0.180\gevc 62\% of the time. The corresponding fraction for other pions in the signal \Bz decay is 5\%. There are differences between the reconstructed $3\pi$ invariant-mass spectrum for the data and that obtained from MC-simulated events. We studied the signal yield before and after reweighting the $3\pi$ invariant-mass spectrum in the MC-simulated events to match the data. The observed change due to the reweighting of the $3\pi$ mass distribution is 1.7\%, which we assign as the associated systematic uncertainty. This also accounts for uncertainties in the relative contributions of the different decay modes and the mass and width of the $a_1^+$ resonance. We use the \Dstarm and \Dzb decay branching fraction uncertainties from Ref.~\cite{ref:pdg}. We use the value of ${\mathcal{B}(\FourS\to\BzBzb) = 0.486\pm0.006}$ from Ref.~\cite{ref:pdg} for the branching fraction of the decay $\FourS \to \BzBzb$, which has a relative uncertainty of 1.2\%. The kaon identification uncertainty is estimated by comparing the number of \Dstarm events in data and MC simulations with and without implementing identification requirements. According to dedicated studies using \babar\ data control samples, we correct for kaon-identification efficiency differences between data and MC simulation by a factor of $0.978\pm0.011$, where the uncertainty is chosen to be half the difference from unity. The signal efficiency MC statistical uncertainty is 0.9\%. Nominally, we subtract the $3\pi$ mass distribution in the sideband from that of the signal region. However, the $3\pi$ mass distribution of both peaking and non-peaking backgrounds in the signal region may not necessarily be the same as that in the sideband. To estimate the associated systematic uncertainty, we test the sideband subtraction procedure using only MC-simulated background events. After applying efficiency corrections to the resulting distribution, we obtain an integral of 571. Dividing this by the number of efficiency-corrected signal in the data, this translates to a 0.7\% difference, which we assign as the associated systematic uncertainty. The number of $B$ mesons produced is uncertain to 0.6\%~\cite{Lees:2013rw}. We study the MC modeling of decay angle correlations, and found the associated systematic uncertainty to be negligible. As described earlier in the text, there is a peaking background contribution in the $m_{\rm ES}$ distribution due to signal events that are misreconstructed. The rate of this background depends on the branching fraction of signal events. Using our measured branching fraction value, we apply corrections to the expected number of \BzBzb peaking background and repeat the signal extraction procedure on the data. There is a small bias on the branching fraction value but it is negligible compared to the systematic uncertainty due to the other peaking backgrounds.

From the number of fitted signal events, corrected for efficiency and normalized to the total number of produced \Bz mesons in the data sample, and taking into account the \Dstarm and \Dzb branching fractions we derive ${\mathcal{B}(\Bz\to\Dstarm\pip\pim\pip) = (7.26 \pm 0.11 \pm 0.31) \times 10^{-3}}$, where the first uncertainty is statistical and the second systematic. The result is consistent with the current world average and is 2.4 times more precise. This result can be used as input for measurements of $\mathcal{R^{(*)}}$ using hadronic $\tau$ decays in the search for deviations from the SM. The inclusive branching fraction value without removing the $D_s^+$ contamination is $(7.37 \pm 0.11 \pm 0.31) \times 10^{-3}$.

We are grateful for the 
extraordinary contributions of our \pep2\ colleagues in
achieving the excellent luminosity and machine conditions
that have made this work possible.
The success of this project also relies critically on the 
expertise and dedication of the computing organizations that 
support \babar.
The collaborating institutions wish to thank 
SLAC for its support and the kind hospitality extended to them. 
This work is supported by the
US Department of Energy
and National Science Foundation, the
Natural Sciences and Engineering Research Council (Canada),
the Commissariat \`a l'Energie Atomique and
Institut National de Physique Nucl\'eaire et de Physique des Particules
(France), the
Bundesministerium f\"ur Bildung und Forschung and
Deutsche Forschungsgemeinschaft
(Germany), the
Istituto Nazionale di Fisica Nucleare (Italy),
the Foundation for Fundamental Research on Matter (The Netherlands),
the Research Council of Norway, the
Ministry of Education and Science of the Russian Federation, 
Ministerio de Econom\'{\i}a y Competitividad (Spain), the
Science and Technology Facilities Council (United Kingdom),
and the Binational Science Foundation (U.S.-Israel).
Individuals have received support from 
the Marie-Curie IEF program (European Union) and the A. P. Sloan Foundation (USA). 



\begin{thebibliography}{99}

\bibitem{ref:babarDstartaunu}
J.P.\ Lees {\em et al.} (\babar\ Collaboration),
Phys.\ Rev.\ D\ {\bf 88}, 072012 (2013).

\bibitem{ref:Belle}
M.\ Huschle {\em et al.} (Belle Collaboration),
Phys.\ Rev.\ D\ {\bf 92}, 072014 (2015).

\bibitem{ref:LHCb}
R.\ Aaij {\em et al.} (LHCb Collaboration),
Phys.\ Rev.\ Lett.\ {\bf 115}, 111803 (2015).

\bibitem{ref:pdg}
K.A.~Olive {\em et al.} (Particle Data Group), 
Chin. Phys. C {\bf 38}, 090001 (2014).

\bibitem{ref:LHCb2}
R.\ Aaij {\em et al.} (LHCb Collaboration),
Phys.\ Rev.\ D\ {\bf 87}, 092001 (2013).

\bibitem{ref:detector}
B.\ Aubert {\em et al.} (\babar\ Collaboration),
Nucl. Instr. Methods Phys. Res. Sect. A {\bf 479}, {1} (2002).

\bibitem{ref:detector2}
B.\ Aubert {\em et al.} (\babar\ Collaboration),
Nucl. Instr. Methods Phys. Res. Sect. A {\bf 729}, 615 (2013).
  
\bibitem{Lees:2013rw} 
J.~P.~Lees {\it et al.}  (\babar\ Collaboration),
Nucl. Instr. Methods Phys. Res. Sect. A {\bf 726}, 203 (2013). 

\bibitem{ref:evtgen}
D.\ J.\ Lange, 
Nucl. Instr. Methods Phys. Res. Sect. A {\bf 462}, 152 (2001). 

\bibitem{ref:geant4}
S.\ Agostinelli {\em et al.} (Geant4 Collaboration), 
Nucl. Instr. Methods Phys. Res. Sect. A {\bf 506}, 250 (2003). 

\bibitem{ref:TMVA}
A.\ H\"{o}cker {\em et al.}, PoS ACAT, 040 (2007), arXiv:physics/0703039.

\bibitem{ref:thrust}
S.\ Brandt  {\em et al.}, Phys.\ Lett.\ {\bf 12}, 57 (1964).

\bibitem{ref:spher}
J.\ Bjorken and S.\ Brodsky, Phys.\ Rev.\ D {\bf  1}, 1416 (1970).

\bibitem{ref:r2all}
G. C. Fox and S. Wolfram, Nucl. Phys. B {\bf  149}, 413 (1979).

\bibitem{ref:crys}
M.\ J.\ Oreglia, Ph.D. thesis, SLAC, Report No. SLAC-R-236, 1980;
J.\ E.\ Gaiser, Ph.D. thesis, SLAC, Report No. SLAC-R-255, 1982;
T.\ Skwarnicki, Ph.D. thesis, INP and DESY, Report No. DESY-F31-86-02, 1986.

\bibitem{ref:argus}
H.\ Albrecht {\em et al.} (ARGUS Collaboration),
Phys.\ Lett.\ B {\bf 241}, 278 (1990).

\bibitem{ref:tracking}
T.\ Allmendinger {\em et al.},
Nucl. Instr. Methods Phys. Res. Sect. A {\bf 704}, 44 (2013).

\end{thebibliography}
\end{document}